%% file: main.tex
\providecommand\IfDocumentMetadataT[1]{}
\providecommand\IfDocumentMetadataF[1]{#1}

\documentclass[sigconf,nonacm]{acmart}
\usepackage{enumitem}
\usepackage{subcaption}
\usepackage{multirow}
\usepackage{tabularx}
\usepackage{threeparttable}
\usepackage{makecell}
\usepackage{xspace}
\usepackage{xurl}
\usepackage{placeins}
\newcolumntype{Y}{>{\raggedright\arraybackslash}X}
\newcolumntype{P}[1]{>{\raggedright\arraybackslash}p{#1}}

\setcopyright{none}
\acmConference{}{}{}
\acmDOI{}
\acmISBN{}
\hypersetup{hidelinks}
\begin{document}
\title[Inherit4Rec]{Inherit4Rec: Parameter Inheritance for Efficient Scaling of Recommendation Models}
\input{authors}
\renewcommand{\shortauthors}{Zhang et al.}

\input{section/abstract}
\maketitle
\raggedbottom

\input{section/introduction}
\input{section/relatedwork}
\input{section/preliminary}
\input{section/methodology}
\input{section/experiment}
\input{section/conclusion}
% \input{section/aiusestatement}
% \input{section/reproducibility}

% Start references after the main paper; the appendix shares the remaining pages.
\clearpage
% Restore normal column output for references and the appendix.
\nobalance
\bibliographystyle{ACM-Reference-Format}
\bibliography{main}

\appendix
\input{section/appendix}

\end{document}

%% file: authors.tex
% Author order, affiliations, emails, and contribution marks are inherited from the source.
% The source does not specify a city for Kuaishou Technology. Add it before final publication.
\author{Ruihao Zhang}
\authornote{Equal contribution.}
\email{ruihaozhang711@gmail.com}
\affiliation{\institution{Kuaishou Technology}\city{Beijing}\country{China}}

\author{Bo Chen}
\authornotemark[1]
\email{renze03@kuaishou.com}
\affiliation{\institution{Kuaishou Technology}\city{Beijing}\country{China}}

\author{Xiao Wang}
\authornotemark[1]
\email{wangxiao35@kuaishou.com}
\affiliation{\institution{Kuaishou Technology}\city{Beijing}\country{China}}

\author{Jinlong Jiao}

\email{jiaojinlong@kuaishou.com}
\affiliation{\institution{Kuaishou Technology}\city{Beijing}\country{China}}

\author{Tijian Hu}

\email{hutijian@kuaishou.com}
\affiliation{\institution{Kuaishou Technology}\city{Beijing}\country{China}}

\author{Qinglin Jia}

\email{dukang05@kuaishou.com}
\affiliation{\institution{Kuaishou Technology}\city{Beijing}\country{China}}

\author{Xiuqiang He}

\email{he.xiuqiang@gmail.com}
\affiliation{\institution{Shenzhen Technology University}\city{Shenzhen}\country{China}}

\author{Xiangyu Zhao}

\email{xianzhao@cityu.edu.hk}
\affiliation{\institution{City University of Hong Kong}\city{Hong Kong}\country{China}}

\author{Chaoyi Ma}
\authornote{Corresponding authors.}
\email{machaoyi03@kuaishou.com}
\affiliation{\institution{Kuaishou Technology}\city{Beijing}\country{China}}

\author{Ruiming Tang}
\authornotemark[2]
\email{tangruiming@kuaishou.com}
\affiliation{\institution{Kuaishou Technology}\city{Beijing}\country{China}}

\author{Wenwu Ou}

\email{luocheng10@kuaishou.com}
\affiliation{\institution{Kuaishou Technology}\city{Beijing}\country{China}}

%% file: section/abstract.tex
\begin{abstract}

Scaling model capacity has emerged as an effective approach to overcoming performance bottlenecks in industrial recommender systems. However, repeatedly training new models from scratch requires large amounts of data and substantial training resources. Parameter inheritance provides a promising route for both dense model growth and sparse conversion, yet existing methods for LLMs are primarily designed for static corpora and can suffer sharp performance drops under dynamically evolving recommendation data. To address these challenges, we propose Inherit4Rec, a parameter-inheritance framework that supports both Dense-to-Dense (D2D) growth and Dense-to-Sparse (D2S) conversion. Inherit4Rec-D2D combines hybrid growth with asymmetric training to preserve the forward function at expansion and maintain update continuity. Inherit4Rec-D2S constructs SMoE networks through co-activation-aware partitioning with a load-balancing loss, preserving dense-model capabilities while promoting balanced expert activation. Experiments on KuaiRand-1K and an industrial short-video recommendation dataset show that both transformations consistently outperform the SOTA inheritance baselines across all prediction objectives. These results demonstrate the effectiveness of Inherit4Rec for continual capacity expansion and computation-efficient sparse conversion in industrial recommender systems.

\end{abstract}

%% file: section/introduction.tex
\section{Introduction}\label{sec:intro}

\begin{figure}[t]
  \centering
  \includegraphics[width=0.8\linewidth]{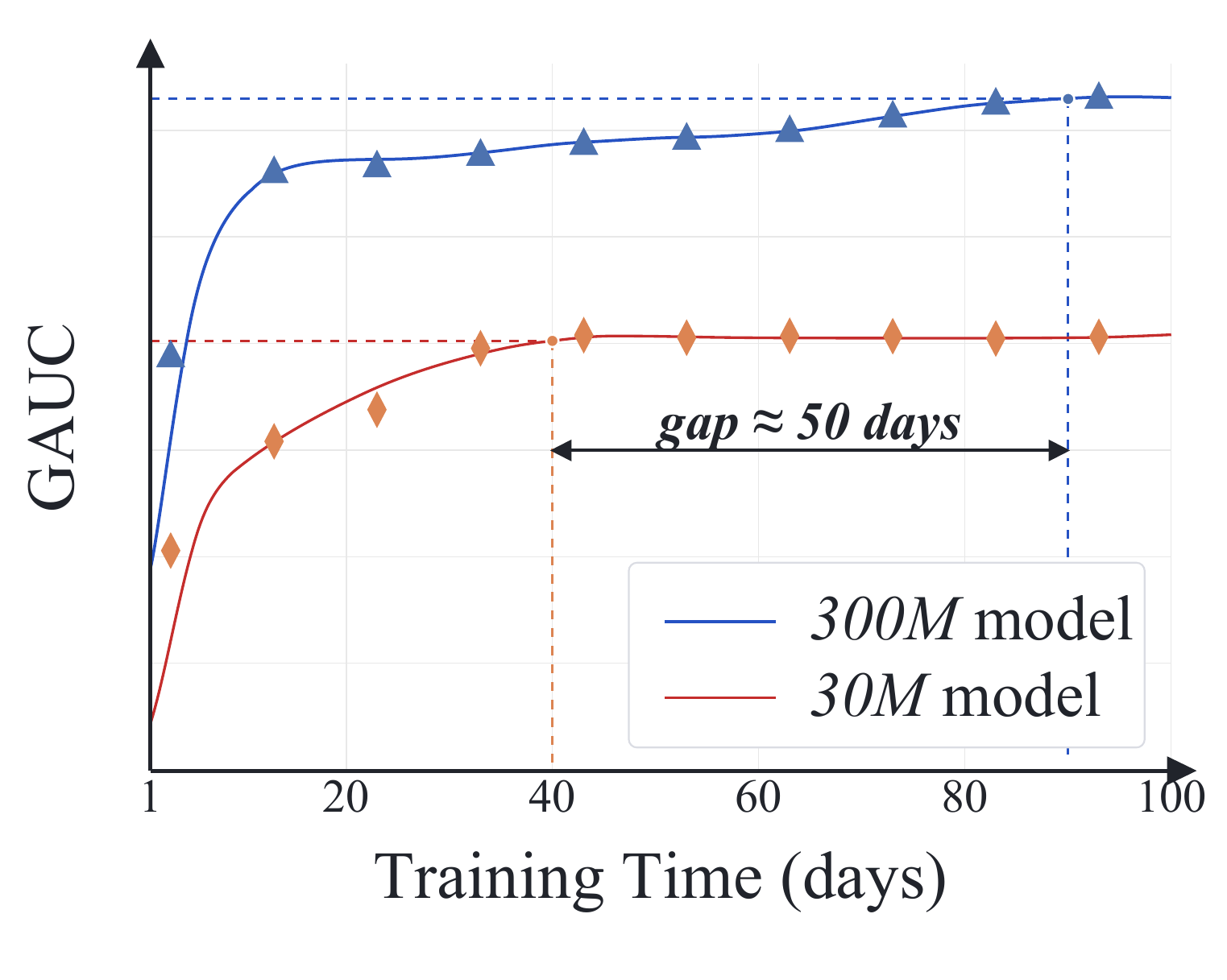}
  \caption{Convergence comparison across models of different parameter scales}
  \Description{Training convergence comparison for recommendation models at different parameter scales.}
  \label{Fig: 300_30}
\end{figure}

Recent advances in recommender systems have demonstrated that increasing model capacity can improve industrial recommendation performance~\citep{hstu,onetrans,li2025survey}. Existing work scales sequence modeling~\citep{longer,stca,sita}, feature interaction~\citep{hhft,rankmixer}, task modeling~\citep{home,smes}, as well as multiple dimensions jointly~\citep{onetrans,hyformer,chen2026uniformer}. To meet the low-latency and high-throughput requirements of industrial serving, TokenMixer-Large further employs a sparse mixture-of-experts (SMoE) architecture for computation-efficient scaling~\citep{tokenmixer}.

Nevertheless, scaling industrial recommender models requires substantially more training data and time, particularly under the common from-scratch paradigm~\citep{tokenmixer, zhang2025dontwaste}. As shown in Figure~\ref{Fig: 300_30}, larger models take considerably longer to converge. Across successive capacity expansions, repeatedly training each larger model from scratch can incur substantial cumulative costs, highlighting the value of reusing previously learned parameters and training states to accelerate convergence.

\begin{figure*}[t]
    \centering

    \begin{subfigure}[t]{0.49\linewidth}
        \centering
        \includegraphics[width=\linewidth]{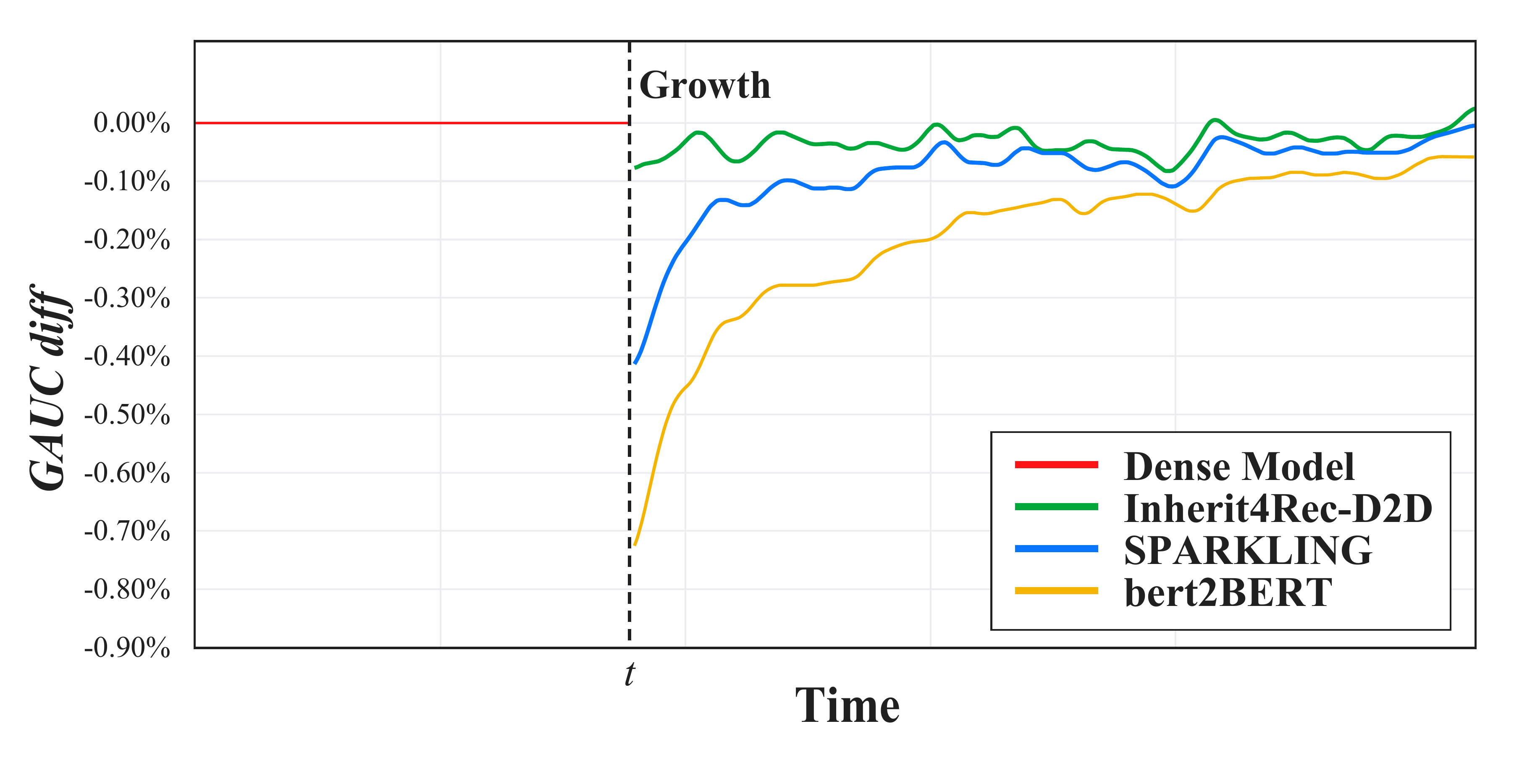}
        \caption{Dense-to-Dense Methods.}
        \label{fig:D2D compare}
    \end{subfigure}
    \hfill
    \begin{subfigure}[t]{0.49\linewidth}
        \centering
        \includegraphics[width=\linewidth]{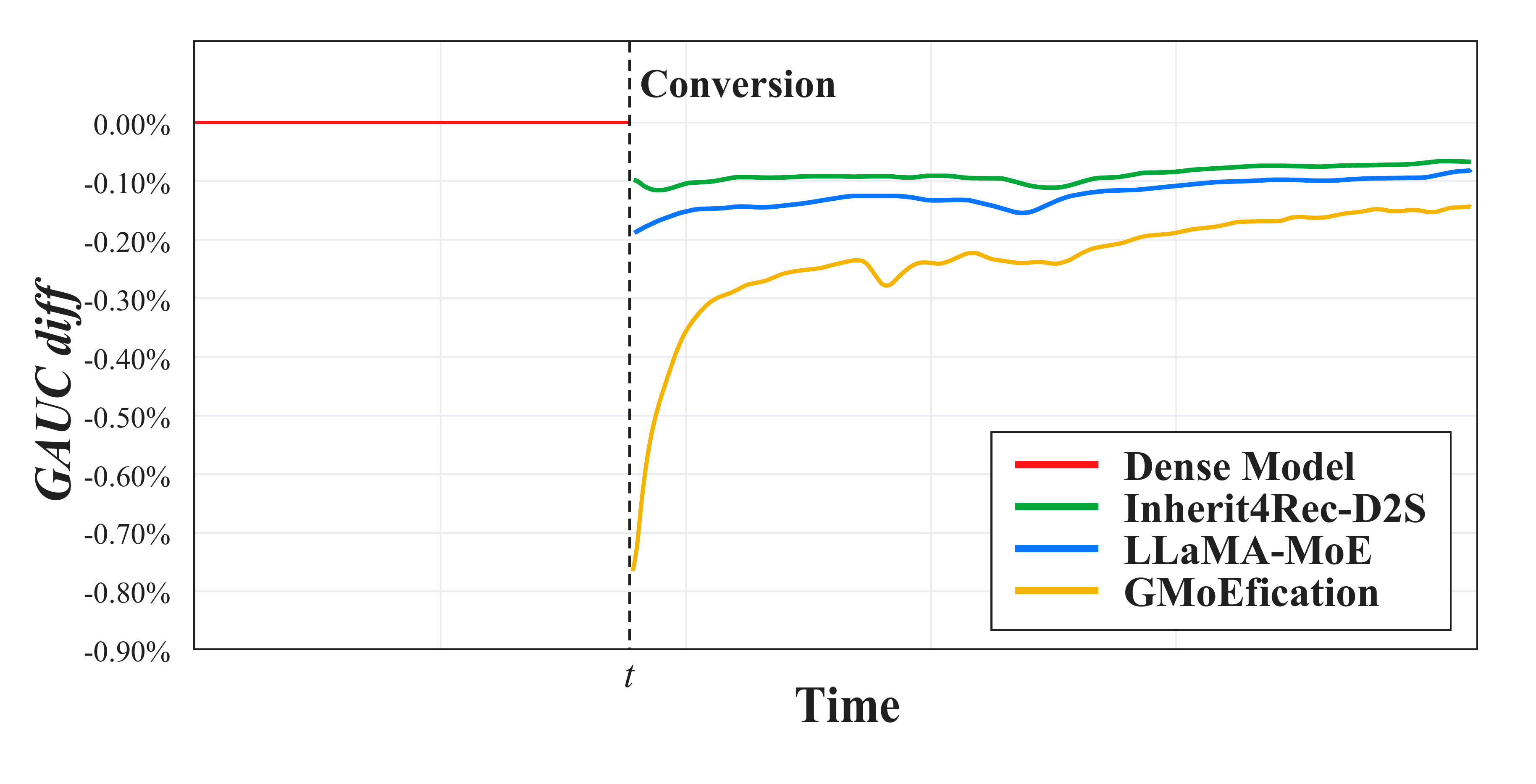}
        \caption{Dense-to-Sparse Methods.}
        \label{fig:D2S compare}
    \end{subfigure}

    \caption{GAUC difference before and after employing the parameter inheritance methods from the large language model domain to industrial streaming recommendation scenarios, where Inherit4Rec is our proposed method.}
    \Description{AUC changes during continued training after Dense-to-Dense growth and Dense-to-Sparse conversion, comparing Inherit4Rec with inheritance baselines.}
    \label{fig:training-curves}

\end{figure*}

\textit{Parameter inheritance} has been extensively studied in LLMs for expanding pretrained dense models~\citep{chen2015net2net, chen2021bert2bert, yu2026sparkling} and converting dense models into MoE architectures for efficient serving~\citep{zhu2024llama, lee2024breaking}. Directly applying these methods to recommender systems is challenging because, unlike language-model pretraining on stable offline corpora, industrial recommenders learn from non-stationary data streams with rapidly evolving user interests, item distributions, and feedback patterns~\cite{rankflow,chan2023capturing}. Consequently, static inheritance cannot keep model parameters aligned with the current data distribution, leading to sharp performance drops after growth or conversion when training resumes, as shown in Figure~\ref{fig:training-curves}.

To address this issue, we propose \textbf{Inherit4Rec}: Parameter Inheritance for Efficient Scaling of Recommendation Models, which achieves efficient and economical scaling up of recommender models through parameter inheritance. It comprises two complementary methods for parameter inheritance: Inherit4Rec-D2D reuses pretrained parameters for Dense-to-Dense (D2D) growth, expanding dense model capacity, whereas Inherit4Rec-D2S reuses them for Dense-to-Sparse (D2S) conversion, turning a large dense model into a sparse one.

Specifically, Inherit4Rec-D2D combines hybrid growth with asymmetric training to support \textit{Forward Preservation} and \textit{Update Continuity}. It expands FFNs by copying their input projections and initializing new output projections so that their contributions sum to zero at expansion, preserving the original output. Inherited parameters retain their optimizer states and learning-rate schedule, while new parameters start with fresh states and a separate schedule. Inherit4Rec-D2S targets \textit{Capability Preservation} through expert partitioning guided by real training samples. Channels with the largest output contributions form an always-on shared expert, while remaining channels that contribute strongly to the same inputs are grouped into equally sized routed experts. The resulting model uses shared-plus-top-1 activation. The experts also inherit the dense model's parameters and optimizer states, and a standard load-balancing loss supports \textit{Balanced Activation} during continued training.

Our contributions can be summarized as follows.
\begin{itemize}[leftmargin=*]
    \item To the best of our knowledge, we are the first to enable efficient continual scaling of recommender models, avoiding costly retraining from scratch and supporting continual capacity expansion and efficient sparsification in industrial systems.
    \item We propose the Efficient Parameter Inheritance Framework (Inherit4Rec), a general framework that leverages parameter inheritance to effectively reduce the resource overhead of Dense-to-Dense growth and Dense-to-Sparse conversion for recommender models.
    \item We combine hybrid growth and asymmetric training in Inherit4Rec-D2D to preserve the forward function and maintain update continuity, and develop expert partitioning based on channel contributions and co-activation statistics from training samples in Inherit4Rec-D2S to preserve dense-model capabilities under sparse activation.
    \item Experiments on KuaiRand-1K and an industrial dataset demonstrate consistent improvements over the evaluated inheritance baselines. On the industrial dataset, Inherit4Rec-D2D expands a 0.31B-parameter model to 1B parameters, achieving a significant GAUC improvement over continued training of the original model. Meanwhile, Inherit4Rec-D2S incurs only marginal GAUC degradation relative to Dense and SMoE models trained from scratch. By reusing existing checkpoints, our method saves 72\% of GPU training time compared with training the target models from scratch.
\end{itemize}

%% file: section/relatedwork.tex
\section{Related Work}\label{sec:related_work}

\begingroup
\setlength{\emergencystretch}{2em}

\subsection{Scalable Recommendation Models}

The empirical success of scaling large language models has motivated similar efforts in industrial recommendation, where scaling feature interaction, sequence modeling, and multi-task learning has improved model capacity and performance.
These efforts explore how additional capacity can be allocated across modeling components while maintaining computational efficiency under the constraints of industrial deployment.
DHEN~\citep{dhen} and Wukong~\citep{wukong} scale hierarchical feature interactions, while HSTU~\citep{hstu} scales sequence modeling.
RankMixer~\citep{rankmixer} combines hardware-friendly token mixing, per-token FFNs, and sparse MoE to build computationally efficient billion-parameter ranking models.
Alongside these developments, OneTrans~\citep{onetrans}, HyFormer~\citep{hyformer}, MixFormer~\citep{mixformer}, and EST~\citep{est} jointly model behavior sequences and non-sequential features in a unified backbone, bringing sequence modeling and feature interaction into a shared framework.
UniFormer~\citep{chen2026uniformer} further extends this paradigm to feature and task spaces through shared interaction structures and multi-view FFNs, broadening the scope of unified scaling to encompass multi-task recommendation.
\subsection{Structural Transformation of Trained Models}

As recommender models and interaction data scale, structural transformation reuses trained checkpoints to reduce the cost of training larger models from scratch. Two directions are central: Dense-to-Dense growth adds parameters to expand capacity, whereas Dense-to-Sparse conversion uses conditionally activated experts to reduce per-example computation; both must control functional changes at conversion and support effective continued training.

\textbf{Dense-to-Dense growth.}
An existing model can be enlarged by leveraging its learned parameters and capabilities. Net2Net~\citep{chen2015net2net} preserves functional equivalence under width or depth expansion, but copy-based initialization can cause gradient symmetry and does not transfer optimizer states, potentially limiting new-parameter learning. For Transformers, bert2BERT~\citep{chen2021bert2bert} initializes larger models from smaller BERT or GPT checkpoints, LiGO~\citep{wang2023learning} learns transfer operators, MSG~\citep{yao2023masked} preserves the original function through progressive masking, and LEMON~\citep{wang2024lemon} enables lossless width and depth expansion. SPARKLING~\citep{yu2026sparkling} combines signal-scale preservation with asymmetric optimization to stabilize expansion and differentiate new parameters. Together, these methods extend model growth from function preservation to stable transformation and effective optimization.

\textbf{Dense-to-Sparse conversion.}
MoE models activate only a subset of experts to increase capacity without proportionally increasing per-input computation, as demonstrated by DeepSeekMoE~\citep{dai2024deepseekmoe} in language modeling and TokenMixer-Large~\citep{tokenmixer} in industrial ranking. Rather than training MoE from scratch, Dense-to-Sparse conversion reuses FFN parameters from a dense model. LLaMA-MoE~\citep{zhu2024llama} constructs experts from LLaMA and continues pretraining, while LLaMA-MoE v2~\citep{qu2024llama} studies expert granularity, activation ratios, and post-conversion training. G-MoEfication~\citep{lee2024breaking} uses balanced K-means and representative activation values to enable sparse execution in models with smooth activation functions such as GeLU and SiLU.

\par\endgroup

%% file: section/preliminary.tex
% !TEX root = ../main.tex

\section{Preliminaries}

\subsection{Streaming Industrial Recommendation}
\label{sec:streaming_recommendation}

Industrial recommender systems are incrementally trained on streaming
interactions~\cite{edcn,ctrl}. At each time window \(t\in\{1,\ldots,T\}\), the model receives
a batch $\mathcal{B}_t
    =
    \left\{
        \left(
            \mathbf{x}_{t,j},
            \mathbf{y}_{t,j}
        \right)
    \right\}_{j=1}^{n_t}$,
where \(\mathbf{x}_{t,j}\) contains user, item, contextual, and sequential
features, and \(\mathbf{y}_{t,j}\) denotes user feedback, such as clicks,
views, and conversions~\cite{autodis}. We denote the complete training checkpoint at time window \(t\) by
\begin{equation}
    \mathcal{C}_t
    =
    \left(
        \boldsymbol{\theta}_t,
        \mathbf{s}_t
    \right),
\end{equation}
where \(\boldsymbol{\theta}_t\) denotes the model parameters and
\(\mathbf{s}_t\) denotes the associated training state, including the
optimizer statistics, learning-rate schedule, and training progress.
The checkpoint is updated as
\begin{equation}
    \mathcal{C}_{t+1}
    =
    \mathcal{U}
    \left(
        \mathcal{C}_t;
        \mathcal{B}_{t+1}
    \right),
\end{equation}
where \(\mathcal{U}\) denotes the update operator.

\subsection{FFN Structural Transformations}
FFNs contain most of the dense parameters in our backbones and are central
to model capacity. We therefore consider two structural transformations of
a pretrained dense FFN.
Let $\Theta_t^{\mathrm{F}}$ denote the parameter set of the pretrained
dense FFN.

\paragraph{Dense-to-Dense (D2D)}
To increase model capacity, D2D expands the original FFN parameter set:
\begin{equation}
    \Theta_t^{\mathrm{F}}
    \xrightarrow{\mathbf{g}^{d}}
    \Theta_t^{\mathrm{d}}
    =
    \Theta_t^{\mathrm{F}}
    \cup
    \Theta_{t,\mathrm{new}}^{\mathrm{d}},
\end{equation}
where $\mathbf{g}^{d}$ is the growth operator and
$\Theta_{t,\mathrm{new}}^{\mathrm{d}}$ denotes the newly introduced
parameters. Thus, the expanded FFN retains all original parameters
while adding new capacity.

\paragraph{Dense-to-Sparse (D2S)}
To reduce per-example computation, D2S reorganizes the dense FFN into a
sparse mixture-of-experts (SMoE), partitioning its parameters into $M$
non-overlapping expert sets:
\begin{equation}
    \Theta_t^{\mathrm{F}}
    \xrightarrow{\mathbf{g}^{m}}
    \left\{
        \Theta_{t,1},
        \ldots,
        \Theta_{t,M}
    \right\},
    \qquad
    \bigcup_{m=1}^{M}
    \Theta_{t,m}
    =
    \Theta_t^{\mathrm{F}},
\end{equation}
where $\mathbf{g}^{m}$ is the partition operator and
$\Theta_{t,m}$ denotes the parameters assigned to the
$m$-th expert. Only a subset of experts is activated for each input.
Router parameters are introduced
separately and are not included in this partition.

\section{Motivation and Design Principles}

% Continual distribution shifts in recommender systems impose additional requirements on parameter inheritance.

\subsection{Dynamic Data Distribution in Recommender Systems}

Industrial recommender systems are continually updated from chronologically
arriving interaction streams~\cite{chan2023capturing,zhang2022towards}. For each time window \(t\), let \(\mathbf{X}\)
denote the user, item, and contextual features, and let \(\mathbf{Y}\) denote
the corresponding supervision signals. Their joint distribution can be
factorized as
\begin{equation}
P_t(\mathbf{X},\mathbf{Y})
=
P_t(\mathbf{X})P_t(\mathbf{Y}\mid\mathbf{X}).
\end{equation}
Changes in user populations, candidate items, and exposure policies may shift
the marginal distribution \(P_t(\mathbf{X})\), while evolving user interests
and feedback patterns may shift the conditional distribution
\(P_t(\mathbf{Y}\mid\mathbf{X})\). Consequently,
\begin{equation}
P_t(\mathbf{X},\mathbf{Y})
\not\approx
P_{t+\Delta}(\mathbf{X},\mathbf{Y}),
\end{equation}
where \(\Delta\) denotes the interval between two data windows. Compared with
the relatively stable offline corpora used in language-model pretraining,
recommendation streams continually shift. Transformations guided primarily
by pretrained parameters can therefore lose performance on these streams
(Figure~\ref{fig:training-curves}). Parameter inheritance should thus retain
learned capabilities while enabling smooth adaptation to subsequent data.

\subsection{Principles of Parameter Inheritance}

To accommodate evolving recommendation data, we formulate the following
principles for D2D growth and D2S conversion. Their formal definitions are provided in
Appendix~\ref{app:additional-evaluation-metrics}.

\noindent\textbf{D2D growth.}
Since the pretrained model has adapted to the current data distribution, D2D
growth should satisfy:
\begin{itemize}[leftmargin=*]

    \item \textit{Forward Preservation}: 
    ensuring identical outputs from the
    expanded and original FFNs for any input before continued training.
    \item \textit{Update Continuity}: ensuring smooth and stable optimization
    trajectories for the inherited channels during subsequent training.
\end{itemize}
\noindent
% Existing D2D methods~\citep{chen2021bert2bert,yu2026sparkling} primarily
% preserve the model function at expansion, but exhibit larger trajectory
% deviations and fluctuations during continued training
% (Section~\ref{sec:exp-visualization}). Effective D2D growth should therefore
% preserve both the current function and its subsequent optimization continuity.

\medskip
\noindent\textbf{D2S conversion.}
As input and activation distributions evolve, D2S conversion should satisfy:

\begin{itemize}[leftmargin=*]
    \item \textit{Capability Preservation}: 
    approximating the original dense
    FFN's outputs as closely as possible under sparse routing.
    \item \textit{Balanced Activation}: balancing the aggregate routing
    load across experts.
\end{itemize}
\noindent
% Existing D2S methods~\citep{zhu2024llama,lee2024breaking} emphasize balanced
% expert construction through parameter-based partitioning, but exhibit high
% output NMSE relative to the original dense FFN
% (Section~\ref{sec:exp-visualization}). This shows that balanced activation
% alone does not guarantee capability preservation, especially when evolving
% activation patterns become misaligned with a static partition.

%% file: section/methodology.tex
% !TEX root = ../main.tex

\section{Methodology}
\label{sec:method}

\begin{figure*}[t]
  \centering
  \includegraphics[width=\textwidth]{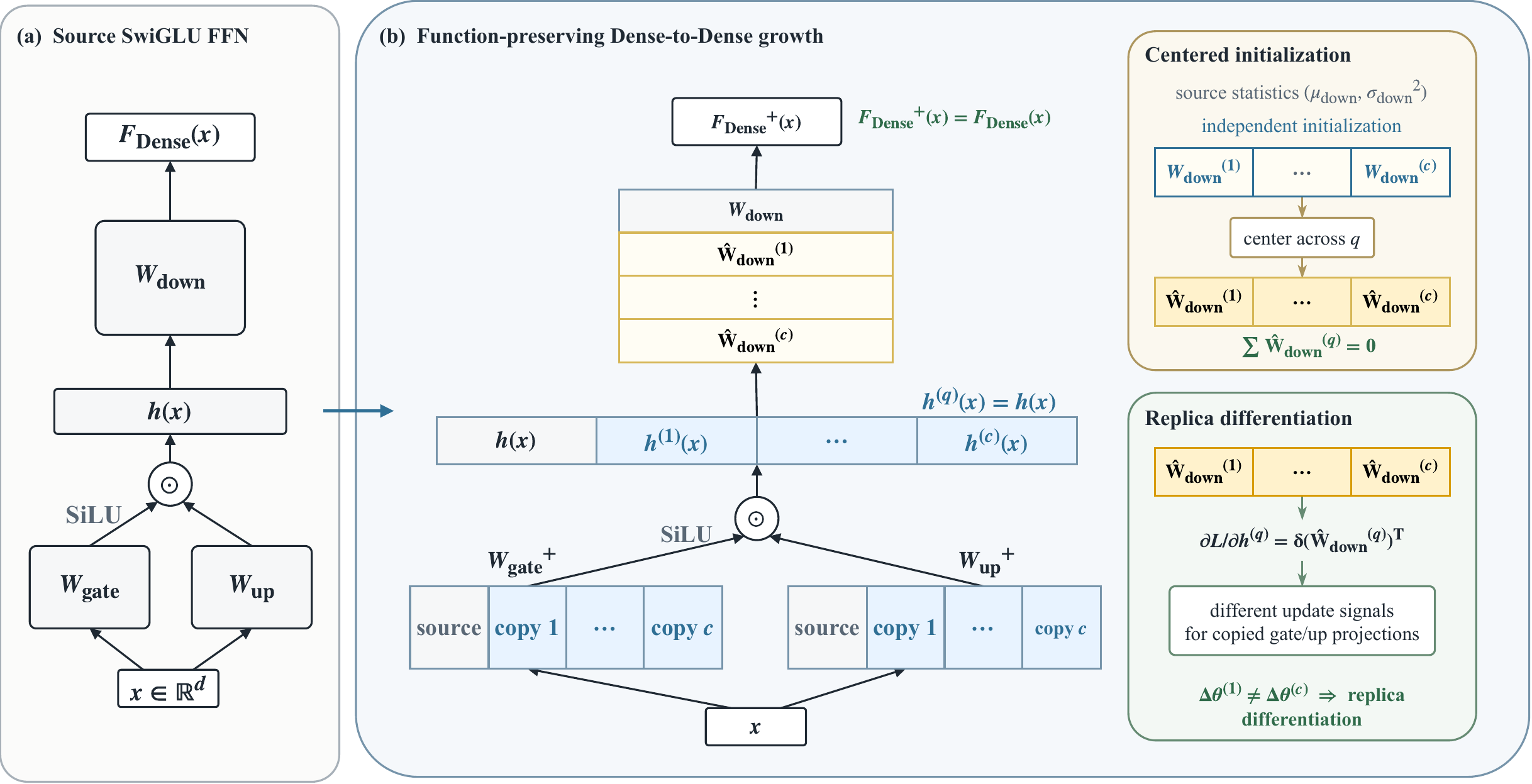}
  \caption{Function-preserving, symmetry-breaking Dense-to-Dense growth.
(a) Source SwiGLU FFN.
(b) Function-preserving Dense-to-Dense growth. Gate and up projections are replicated, while new down projections
are sampled from the source $\mathbf{W_{\mathrm{down}}}$ statistics and zero-sum
centered, preserving the initial output and enabling replica
differentiation during continued training.}
  \Description{SwiGLU growth diagram showing copied gate and up projections and zero-sum centered new down projections.}
    \label{fig:dense-to-dense-growth}
\end{figure*}

\subsection{Overview of FFN Transformation}
\label{sec:method-overview}

% Starting from the streaming checkpoint $\mathcal C_t$ defined in
% Section~\ref{sec:streaming_recommendation}, we construct a transformed
% checkpoint $\mathcal C_t^{+}$ from which training continues without a restart.
% We focus on the model’s per-token FFNs, whose channel structure provides a direct way to expand capacity or reduce active computation. 
% We transform these FFNs and map their associated training state, leaving all other modules unchanged at conversion.  

For a SwiGLU FFN with input/output dimension $d$ and hidden dimension
$d_{\mathrm{hid}}$, the hidden activations and output for an input
$\mathbf{x}\in\mathbb{R}^{d}$ are
\begin{equation}
\begin{gathered}
\mathbf{h}(\mathbf{x})
= \operatorname{SiLU}(\mathbf{x}\mathbf{W_{\mathrm{gate}}})
\odot (\mathbf{x}\mathbf{W_{\mathrm{up}}}),\\
\mathbf{F}_{\mathrm{Dense}}(\mathbf{x})
= \mathbf{h}(\mathbf{x})\mathbf{W_{\mathrm{down}}},
\end{gathered}
\label{eq:channel-ffn}
\end{equation}
where $\mathbf{W_{\mathrm{gate}}},\mathbf{W_{\mathrm{up}}}\in
\mathbb{R}^{d\times d_{\mathrm{hid}}}$,
$\mathbf{W_{\mathrm{down}}}\in\mathbb{R}^{d_{\mathrm{hid}}\times d}$,
and $\mathbf{h}(\mathbf{x})\in\mathbb{R}^{d_{\mathrm{hid}}}$.

Let $h_j(\mathbf{x})$ denote the $j$-th hidden activation and
$[\mathbf{W}_{\mathrm{down}}]_{j,:}$ the $j$-th row of $\mathbf{W_{\mathrm{down}}}$.
The output then decomposes into hidden-channel contributions:
\begin{equation}
\mathbf{F}_{\mathrm{Dense}}(\mathbf{x})
= \sum_{j=1}^{d_{\mathrm{hid}}}
h_j(\mathbf{x})[\mathbf{W}_{\mathrm{down}}]_{j,:}.
\label{eq:channel-ffn-decomposition}
\end{equation}

Each hidden activation $h_j(\mathbf{x})$ is computed from the $j$-th
columns of $\mathbf{W_{\mathrm{gate}}}$ and $\mathbf{W_{\mathrm{up}}}$, and contributes to
the output through the $j$-th row of $\mathbf{W_{\mathrm{down}}}$. We therefore
treat these three aligned parameter slices as a single channel-level
unit in both transformations.

% Queue this two-column float early so it appears on page 5.
\begin{figure*}[t]
  \centering
  \includegraphics[width=\textwidth]{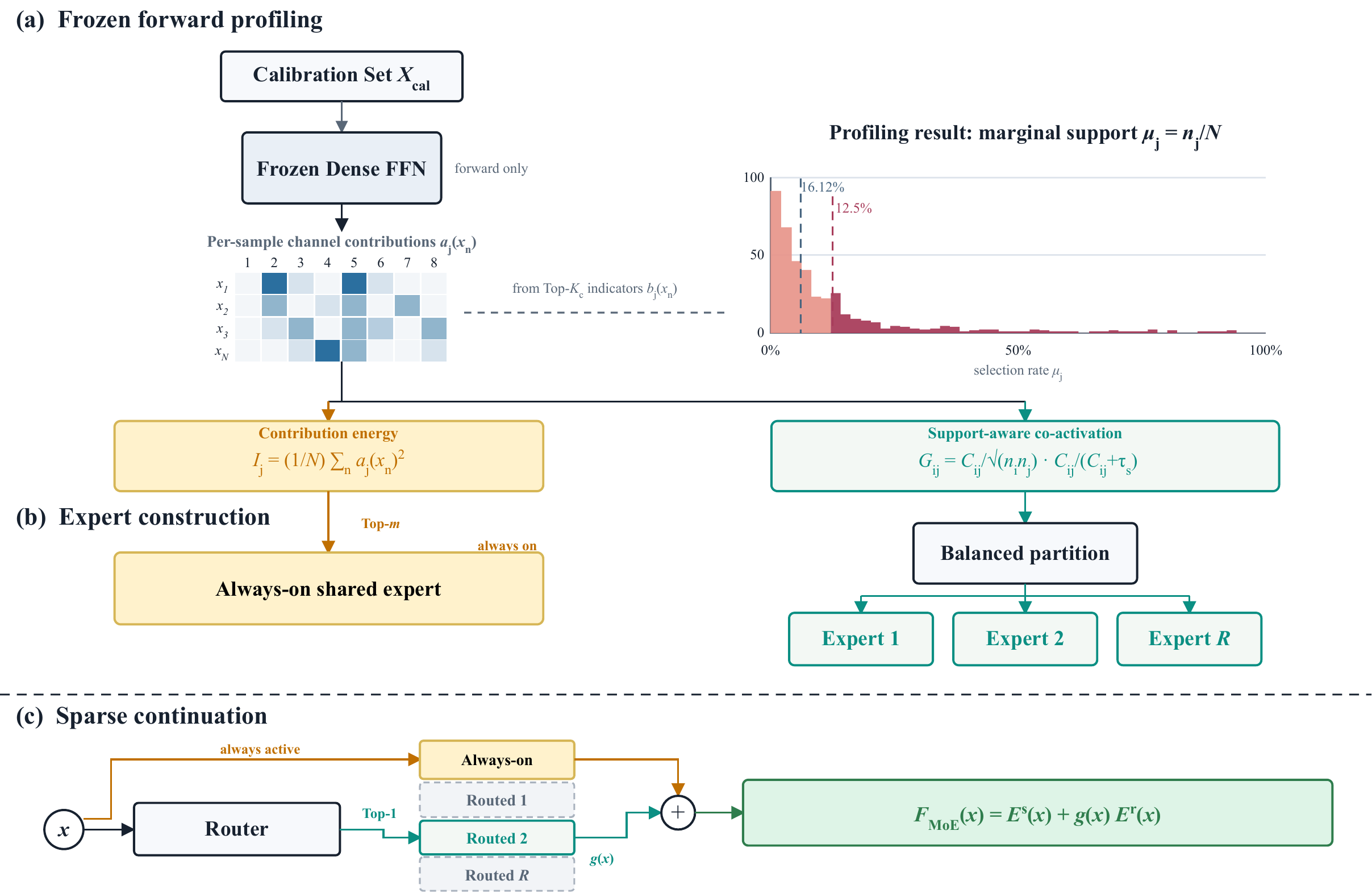}
  \caption{Contribution-aware Dense-to-Sparse conversion. A forward-only
  pass ranks channels by empirical output contribution.  The
  top-ranked channels form an always-on shared expert, while the remaining
  channels are divided into routed experts using a balanced partition based on
  contribution co-activation.  All experts reuse channel-aligned dense
  parameters and optimizer states. }
  \Description{Conversion pipeline selecting high-contribution shared channels and grouping remaining channels into routed experts using co-activation statistics.}
    \label{fig:dense-to-sparse-conversion}
\end{figure*}

\subsection{Dense-to-Dense Growth}
\label{sec:dense-growth-method}

Inherit4Rec-D2D expands FFN capacity on evolving data streams under two
principles: \emph{forward preservation} and \emph{update continuity}. Specifically, it combines parameter
replication with zero-sum random initialization to preserve the pre-growth
function, and applies asymmetric optimization to inherited and new parameters
to support smooth continued training.

\paragraph{Hybrid Growth Strategy.}
To preserve the forward function while enabling new channels to
differentiate, we replicate the gate and up projections and randomly
initialize the new down projections under a zero-sum constraint.
As illustrated in Figure~\ref{fig:dense-to-dense-growth}, we add $c$
replicas of each channel in a source FFN with hidden dimension
$d_{\mathrm{hid}}$, increasing its width to
$d_{\mathrm{hid}}^{+}=(c+1)d_{\mathrm{hid}}$, where $c\geq 1$.
For the $q$-th replica block, we set
\begin{equation}
\begin{aligned}
\mathbf{W_{\mathrm{gate}}^{(q)}}&=\mathbf{W_{\mathrm{gate}}},\\
\mathbf{W_{\mathrm{up}}^{(q)}}&=\mathbf{W_{\mathrm{up}}},
\qquad q=1,\ldots,c.
\end{aligned}
\label{eq:d2d-copy-input-projections}
\end{equation}
All new blocks therefore have identical hidden activations immediately
after growth:
$\mathbf h^{(q)}(\mathbf x)=\mathbf h(\mathbf x)$.

Let $\mu_{\mathrm{down}}$ and $\sigma_{\mathrm{down}}^2$ denote the
empirical entrywise mean and variance of the source
$\mathbf{W_{\mathrm{down}}}$. We independently sample the entries of each new
down-projection block
$\mathbf{W_{\mathrm{down}}^{(q)}}\in\mathbb R^{d_{\mathrm{hid}}\times d}$
from $\mathcal N(\mu_{\mathrm{down}},\sigma_{\mathrm{down}}^2)$
and center the blocks across replicas:
\begin{equation}
\begin{gathered}
\mathbf{\widehat W_{\mathrm{down}}^{(q)}}
=
\mathbf{W_{\mathrm{down}}^{(q)}}
-\frac{1}{c}\sum_{r=1}^{c}\mathbf{W_{\mathrm{down}}^{(r)}},
\\
\sum_{q=1}^{c}\mathbf{\widehat W_{\mathrm{down}}^{(q)}}=\mathbf 0.
\end{gathered}
\label{eq:d2d-group-centering}
\end{equation}
%The source projections remain unchanged. 
Since the new blocks have
identical initial activations, their output contributions cancel:
\begin{equation}
\begin{aligned}
\mathbf{F}_{\mathrm{Dense}}^{+}(\mathbf x)
&=\mathbf{F}_{\mathrm{Dense}}(\mathbf x)
+\sum_{q=1}^{c}\mathbf h^{(q)}(\mathbf x)
\mathbf{\widehat W_{\mathrm{down}}^{(q)}}\\
&=\mathbf{F}_{\mathrm{Dense}}(\mathbf x)
+\mathbf h(\mathbf x)\sum_{q=1}^{c}
\mathbf{\widehat W_{\mathrm{down}}^{(q)}}\\
&=\mathbf{F}_{\mathrm{Dense}}(\mathbf x).
\end{aligned}
\label{eq:d2d-function-preservation}
\end{equation}
Thus, hybrid growth ensures forward stability by exactly preserving
the source FFN function at expansion in exact arithmetic.

The centered down projections are generally distinct,
allowing the replicated channels to receive different gradient signals.
Specifically, given the row-vector upstream gradient
$\boldsymbol\delta$ at the FFN output,
\begin{equation}
\frac{\partial\mathcal L}{\partial\mathbf h^{(q)}}
=
\boldsymbol\delta
\bigl(\mathbf{\widehat W_{\mathrm{down}}^{(q)}}\bigr)^\top.
\label{eq:d2d-replica-gradient}
\end{equation}
Under non-degenerate gradients, these differences enable the copied
gate and up projections to develop distinct representations during
continued training. The zero-sum constraint is applied only at
initialization.

\paragraph{Asymmetric Training Strategy.}
To preserve update continuity while allowing the added capacity to adapt,
inherited parameters retain their first- and second-order optimizer moments
and existing learning-rate schedule, whereas new parameters start with zero
moments and follow a separate warmup-and-decay schedule. This includes the
new gate and up projections: copying their parameter values does not copy
their optimizer states.
Appendix~\ref{app:state-migration} details the state mapping.

\subsection{Dense-to-Sparse Conversion}
\label{sec:dense-to-sparse-method}

To retain the dense FFN's learned capabilities as much as possible,
Inherit4Rec-D2S reorganizes its channels into a sparse MoE using forward-pass
statistics collected at conversion: consistently high-contribution channels
form an always-on shared expert, while the remaining channels are partitioned
into routed experts according to contribution co-activation.

\paragraph{Co-activation-aware Partitioning.}
As illustrated in Figure~\ref{fig:dense-to-sparse-conversion}, we partition
the $d_{\mathrm{hid}}$ channels of the source Dense FFN into one shared
expert and $R$ routed experts with equal width 
$m=d_{\mathrm{hid}}/(R+1)$. Each input activates the shared expert and
one routed expert, so only $2m$ channels are active, corresponding to
$2/(R+1)$ of the original FFN width.

\textit{Channel allocation.}
To select the shared channels and group the remaining channels into routed
experts, we freeze the Dense checkpoint and, for each FFN, run a forward-only
pass on $N$ samples $\{\mathbf{x}_n\}_{n=1}^{N}$ from the stream available
at conversion. We score channel $j$ by its mean squared
contribution to the Dense FFN output:
\begin{equation}
\begin{aligned}
I_j&=\frac{1}{N}\sum_{n=1}^{N}
\left\|h_j(\mathbf x_n)[\mathbf{W}_{\mathrm{down}}]_{j,:}\right\|_2^2,
\\[-2pt]
&\hspace{2em}j=1,\ldots,d_{\mathrm{hid}}.
\end{aligned}
\label{eq:d2s-contribution-energy}
\end{equation}

% Queue the first experimental table on page 5 for placement on page 6.

The top-$m$ channels form the always-on shared expert. For the remaining
channels, we define a pairwise similarity $G_{ij}$ from the co-occurrence of
large output contributions across samples, correcting for individual channel
frequencies and downweighting low-support pairs. Let $E_r$ denote the channel
set of routed expert $r$. We partition the remaining channels into $R$
disjoint experts with width $m$ by maximizing within-expert similarity:
\begin{equation}
\begin{aligned}
\max_{\{E_r\}_{r=1}^{R}}\quad
& \sum_{r=1}^{R}\sum_{\substack{i<j\\i,j\in E_r}}G_{ij},\\
\text{s.t.}\quad
& |E_r|=m,\qquad
E_r\cap E_{r'}=\varnothing\quad(r\ne r').
\end{aligned}
\label{eq:d2s-allocation-objective}
\end{equation}
This groups channels that tend to contribute strongly to the same inputs,
allowing them to be evaluated together when their expert is selected. The
definition of $G_{ij}$ and the partitioning algorithm are given in
Appendix~\ref{app:d2s-partition-algorithm}.

\paragraph{Sparse Execution and State Inheritance.}
For an input $\mathbf x$, a bias-free router
$\mathbf{W}_R\in\mathbb R^{d\times R}$ produces routing probabilities and selects
one routed expert:
\begin{equation}
\begin{aligned}
\mathbf p(\mathbf x)&=\operatorname{softmax}(\mathbf x\mathbf{W}_R),\\
r^*(\mathbf x)&=\arg\max_r p_r(\mathbf x),\qquad
 g(\mathbf x)=p_{r^*(\mathbf x)}(\mathbf x).
\end{aligned}
\label{eq:d2s-selected-gate}
\end{equation}
Let $\mathbf{E}^{\mathrm s}(\mathbf x)$ and $\mathbf{E}^{\mathrm r}(\mathbf x)$
denote the outputs of the shared expert and the selected routed expert,
respectively. The sparse FFN output is
\begin{equation}
\mathbf{F}_{\mathrm{MoE}}(\mathbf x)
=\mathbf{E}^{\mathrm s}(\mathbf x)+g(\mathbf x)\mathbf{E}^{\mathrm r}(\mathbf x).
\label{eq:d2s-sparse-output}
\end{equation}
Each expert inherits channel-aligned gate, up, and down parameter slices,
together with the corresponding optimizer-state slices.
Appendix~\ref{app:state-migration} gives the exact slicing and router initialization.

\paragraph{Load-balanced Continued Training.}
Let $f_r$ be the fraction of samples in a batch assigned to expert $r$,
and $\bar p_r$ its mean routing probability before top-1 selection. We use the
Switch-style auxiliary loss~\citep{fedus2021switch} to enable load-balanced continued training:
\begin{equation}
\mathcal L_{\mathrm{bal}}=R\sum_{r=1}^{R}f_r\bar p_r.
\label{eq:d2s-balance-loss}
\end{equation}
The losses are averaged over transformed FFNs and combined with the task loss:
\begin{equation}
\mathcal L=\mathcal L_{\mathrm{task}}
+\lambda_{\mathrm{bal}}\mathcal L_{\mathrm{bal}}.
\label{eq:d2s-training-objective}
\end{equation}
For reproducibility, we provide the hyperparameters for forward-pass statistics
collection, channel partitioning, and continued training in
Appendix~\ref{app:method-hyperparameters}.

%% file: section/experiment.tex
% !TEX root = ../main.tex
% Allow wide floats to share text pages during the experiment section.
\begingroup
\renewcommand{\dbltopfraction}{0.85}
\raggedbottom

\section{Experiments}
\label{sec:exp}

\subsection{Experimental Settings}
\label{sec:exp-setup}

\paragraph{Datasets and Metrics.}
We evaluate on KuaiRand-1K~\citep{gao2022kuairand} and an industrial
short-video recommendation dataset. Each dataset includes four prediction
tasks covering viewing behavior and explicit user engagement. %, as listed inTables~\ref{tab:d2d-main} and~\ref{tab:d2s-main}. 
We report GAUC for each
task and use UniFormer~\citep{chen2026uniformer} as the default backbone due to well-balanced parameter allocation.

\paragraph{Baselines and Comparison Protocol.}
For D2D growth, we compare against bert2BERT-FPI~\citep{chen2021bert2bert} and
SPARKLING~\citep{yu2026sparkling}, with Seed (Base) and Train-from-Scratch (Oracle)
serving as reference models. In D2D, Seed (Base) continues training the
source model without expansion. For D2S conversion, we compare against
LLaMA-MoE~\citep{zhu2024llama} and G-MoEfication~\citep{lee2024breaking},
with Seed (Base), Dense Train-from-Scratch, and SMoE Train-from-Scratch (Oracle)
as references. In D2S, Seed (Base) is a small Dense reference whose active
parameter count is comparable to that of the converted sparse models.
The source checkpoint used for D2S conversion is a separate pretrained
large Dense model. Dense Train-from-Scratch provides a large dense
reference, while SMoE Train-from-Scratch matches the target sparse architecture.

\paragraph{Training Protocol.}
The industrial dataset contains approximately 50 billion examples per day
on average. For D2D and D2S, source models are trained on the first 18 and
20 daily windows, followed by continued training on seven and eight
subsequent windows after transformation, respectively. The corresponding
Train-from-Scratch references are trained on the complete 25-window and
28-window streams, respectively.
For KuaiRand-1K, D2D and D2S models undergo five and three epochs of
continued training after transformation, respectively.
On both datasets, the D2S Seed (Base) undergoes continued training on the
same data and for the same duration as the converted models.
Incremental training costs are accounted for from an existing Dense
checkpoint. The 72\% saving refers to the industrial D2D training-window
budget, as detailed in Appendix~\ref{app:cost-accounting}.
Dataset construction and full configurations
are provided in Appendix~\ref{app:experimental-details}.

\subsection{Main Results}
\label{sec:exp-main-results}

\input{section/table_d2d}

% Tables~\ref{tab:d2d-main} and~\ref{tab:d2s-main} report the main
% results for Dense-to-Dense growth and Dense-to-Sparse conversion,
% respectively.

% Queue the second wide table ahead of the Dense-to-Sparse discussion.
\begin{table*}[!t]
\caption{Offline Dense-to-Sparse GAUC results. Seed (Base) is a small Dense
reference with an active parameter count comparable to that of the converted
sparse models. Bold and underlined values
indicate the best and second-best results among transformation methods,
respectively.}
\label{tab:d2s-main}
\centering
\footnotesize
\setlength{\tabcolsep}{1.5pt}
\renewcommand{\arraystretch}{1.05}
\begin{threeparttable}
\begin{tabular*}{\linewidth}{
@{\extracolsep{\fill}}
P{0.14\linewidth}
P{0.07\linewidth}
*{8}{c}
@{}}
\toprule
& & \multicolumn{4}{c}{KuaiRand-1K Dataset}
& \multicolumn{4}{c}{Industrial Dataset} \\
\cmidrule(lr){3-6}
\cmidrule(lr){7-10}

\multicolumn{2}{l}{Model}
& Effective-view & Long-view & Comment & Like
& Effective-view & Long-view & Follow & Like \\
\midrule

\multicolumn{2}{l}{Seed (Base)}
& 0.6631 & 0.6987 & 0.6773 & 0.6690
& 0.7557 & 0.7662 & 0.8318 & 0.8335 \\

\multirow{2}{*}{\makecell[l]{Train-from-Scratch}}
& Dense
& 0.6664 & 0.7018 & 0.6950 & 0.6707
& 0.7570 & 0.7675 & 0.8332 & 0.8350 \\

& SMoE 
& 0.6659 & 0.7014 & 0.6947 & 0.6706
& 0.7567 & 0.7674 & 0.8329 & 0.8349 \\

\midrule

\multicolumn{2}{l}{LLaMA-MoE}
& 0.6632 & 0.7006 & \underline{0.6876} & 0.6680
& \underline{0.7562} & \underline{0.7667}
& \underline{0.8326} & \underline{0.8344} \\

\multicolumn{2}{l}{G-MoEfication}
& \underline{0.6651} & \underline{0.7007}
& 0.6821 & \underline{0.6691}
& 0.7561 & \underline{0.7667}
& 0.8325 & 0.8343 \\

\midrule

\multicolumn{2}{l}{Inherit4Rec-D2S}
& \textbf{0.6654} & \textbf{0.7012}
& \textbf{0.6927} & \textbf{0.6701}
& \textbf{0.7566} & \textbf{0.7672}
& \textbf{0.8328} & \textbf{0.8347} \\

\multicolumn{2}{l}{Impr.}
& 0.05\% & 0.07\% & 0.74\% & 0.15\%
& 0.05\% & 0.07\% & 0.02\% & 0.04\% \\

\bottomrule
\end{tabular*}

\begin{tablenotes}[flushleft]
\footnotesize
\item Seed (Base) uses the same continuation data and duration as the
converted models and is distinct from the large Dense checkpoint used
for conversion. Seed and both Train-from-Scratch models are references and are
excluded from ranking and baseline selection; SMoE serves as the
target-architecture Oracle. Impr. follows the definition in
Table~\ref{tab:d2d-main}.
\end{tablenotes}
\end{threeparttable}
\end{table*}

\paragraph{Dense-to-Dense growth.}
As shown in Table~\ref{tab:d2d-main}, Inherit4Rec-D2D outperforms
bert2BERT-FPI and SPARKLING on all four tasks for both datasets. Its mean
relative GAUC gains over the strongest inheritance baseline for each task
are 0.15\% on KuaiRand-1K and 0.05\% on the industrial dataset, averaging
the per-task relative gains over the four tasks in each dataset. Relative
to the non-growing Seed (Base), it achieves mean
relative GAUC gains of 0.76\% and 0.11\% on KuaiRand-1K and the industrial
dataset, respectively. These gains are achieved through post-growth
continuation from an existing checkpoint, at a small fraction of the cost of
training the target model from scratch.

Inherit4Rec-D2D achieves GAUC close to that of the oracle model trained
from scratch. With additional streaming data and longer
post-growth training, the GAUC gap on the industrial
dataset narrows to a negligible level
(Figure~\ref{fig:d2d-extended-training}(a)). These results show that
parameter inheritance enables effective model growth while approaching
the performance of the larger model trained from scratch.

\paragraph{Dense-to-Sparse conversion.}
Under matched FFN activation budgets, Inherit4Rec-D2S outperforms
LLaMA-MoE and G-MoEfication on all four tasks in each dataset
(Table~\ref{tab:d2s-main}).
Its mean relative GAUC gains over the strongest inheritance baseline for
each task are 0.25\% on KuaiRand-1K and 0.04\% on the industrial dataset,
using the same per-task averaging protocol as D2D.

Inherit4Rec-D2S also outperforms the small Dense reference Seed (Base)
on all four tasks in each dataset, with mean relative GAUC gains of 0.79\%
and 0.13\% on KuaiRand-1K and the industrial dataset, respectively.
Both models use the same continuation data and duration, with comparable
active parameter counts. This comparison shows the benefit of reusing a
larger Dense model's parameters through sparse execution relative to
continued training of a small Dense model at a comparable active-parameter budget.

Compared with the oracle model SMoE Train-from-Scratch, Inherit4Rec-D2S has mean relative
GAUC gaps of 0.117\% on KuaiRand-1K and 0.019\% on the industrial dataset.
These results show that converting an existing
large Dense checkpoint can achieve performance close to that of the target
sparse model trained from scratch under the same FFN activation budget.

\subsection{Ablation Studies}
\label{sec:exp-ablation}

\paragraph{Dense-to-Dense components.}

We evaluate the contributions of zero-sum initialization and optimizer-state
transfer to Inherit4Rec-D2D on the industrial dataset, which are
designed to support forward preservation and update continuity, respectively.
The two ablation variants remove zero-sum initialization
(\emph{w/o zero-sum}) and reset all Adam states
(\emph{w/o optimizer transfer}).

\begin{figure}[!htbp]
\centering
\includegraphics[width=\columnwidth]{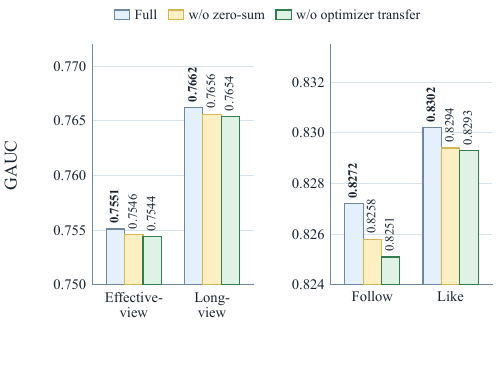}
\captionsetup{skip=3pt}
\caption{Dense-to-Dense ablation.
Full is Inherit4Rec-D2D; best values are bold.
Panels use different vertical scales.}
\label{fig:d2d-ablation}
\Description{Dense-to-Dense grouped-bar ablation results for four industrial tasks in two panels with different GAUC scales. Full outperforms both removing zero-sum initialization and resetting Adam states on every task. cb{a,b}}
\end{figure}

Removing either zero-sum initialization or optimizer-state transfer leads
to clear performance degradation across all four tasks
(Figure~\ref{fig:d2d-ablation}), demonstrating the contribution of both
components to effective model growth. These results highlight their
complementary roles: zero-sum initialization preserves the model's
function at expansion, while optimizer-state transfer retains the learned
optimization history to support stable post-growth training.

\paragraph{Dense-to-Sparse components.}

We evaluate three components of Inherit4Rec-D2S: co-activation-aware
partitioning, shared-plus-top-1 execution, and the load-balancing loss. The
\emph{w/o co-act partition} variant replaces co-activation-aware partitioning
with contiguous partitioning; \emph{w/o shared expert} replaces the shared
expert with a routed expert while keeping the activated parameter count
unchanged; and \emph{w/o balancing} removes the auxiliary load-balancing loss.

\begin{figure}[!htbp]
\centering
\includegraphics[width=\columnwidth]{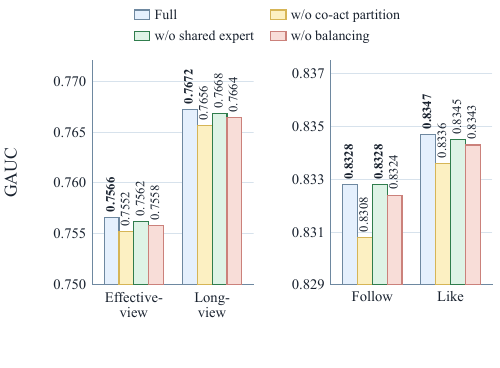}
\captionsetup{skip=3pt}
\caption{Dense-to-Sparse ablation.
Full is Inherit4Rec-D2S; best values, including ties, are bold.
Panels use different vertical scales. }
\label{fig:d2s-ablation}
\Description{Dense-to-Sparse grouped-bar ablation results for four industrial tasks in two panels with different GAUC scales. Replacing co-activation-aware partitioning causes the largest drops. Replacing the shared expert with a routed expert at the same activated parameter count lowers GAUC on three tasks, while removing the balancing loss lowers GAUC on all four tasks.}
\end{figure}

Replacing co-activation-aware partitioning causes the largest performance
decline across all four tasks (Figure~\ref{fig:d2s-ablation}), identifying
it as the most influential component among those evaluated. This suggests
that preserving relationships among channels that contribute together is
central to retaining predictive capabilities under sparse activation.
Replacing the shared expert lowers performance on most tasks, supporting
the value of keeping high-contribution channels consistently active.
Removing the load-balancing loss also degrades performance across all
four tasks, indicating that expert-load regularization contributes to
effective training after conversion. Together, these results support
the complementary roles of expert partitioning, shared computation, and
load balancing in preserving the quality of the converted model.

\subsection{Backbone Compatibility and Scalability}
\label{sec:exp-extended-growth}

\paragraph{Backbone Compatibility.}
To assess compatibility beyond UniFormer, we evaluate Inherit4Rec with
RankMixer~\citep{rankmixer}, a representative architecture for scaling
industrial recommendation models. We follow the same experimental settings
as in the UniFormer experiments.

The results in Table~\ref{tab:rankmixer-compatibility} in the appendix
confirm the effectiveness of both D2D growth and D2S conversion with the
RankMixer backbone.

\paragraph{Extended Post-Growth Training.}
We examine how Inherit4Rec-D2D benefits from data scaling by extending
post-growth training to incorporate more samples from the industrial data
stream. As shown in Figure~\ref{fig:d2d-extended-training}(a), its
Effective-view GAUC gap to Train-from-Scratch (Oracle) continues to narrow
as the cumulative amount of training data increases. This trend suggests
that the expanded model can make effective use of additional data,
translating its increased capacity into further performance gains and
progressively approaching the larger model trained from scratch.

\begin{figure}[!htbp]
\centering
\includegraphics[width=0.90\columnwidth]{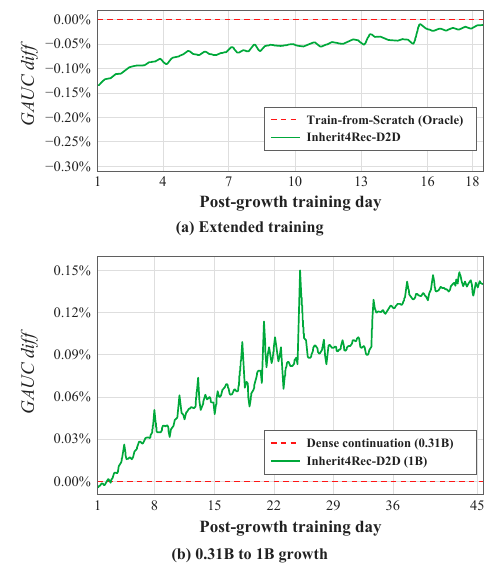}
\captionsetup{skip=3pt}
\caption{Absolute Effective-view GAUC differences (\%).
References: (a) Train-from-Scratch (Oracle); (b) Dense continuation
(0.31B). Panel (b) reports measured GAUC differences over 45 days of
continued training.}
\label{fig:d2d-extended-training}
\Description{Two vertically stacked panels show Effective-view GAUC differences. The upper panel shows extended D2D training approaching the Train-from-Scratch reference. The lower panel shows measured gains from growing a 0.31B model to 1B relative to continued training of the 0.31B model over 45 days. Red dashed zero lines denote equal performance with each panel's reference.}
\end{figure}

\paragraph{Growth to One Billion Parameters.}
To assess growth at a larger scale, we expand a 0.31B-parameter model
pretrained for four months to 1B parameters and continue training both the
expanded model and the original 0.31B model for 45 days. At day 45, the
expanded model achieves an absolute Effective-view GAUC gain of 0.14\%
over continued training of the original model
(Figure~\ref{fig:d2d-extended-training}(b)). This result supports the
effectiveness of Inherit4Rec-D2D for scaling a mature checkpoint to one
billion parameters. Reusing the pretrained checkpoint also substantially
reduces training costs compared with training the larger model from scratch.

\subsection{Empirical Validation of Design Principles}
\label{sec:exp-additional-analyses}

We examine Inherit4Rec with respect to the four design principles.
Formal definitions of the diagnostic metrics are provided in
Appendix~\ref{app:additional-evaluation-metrics}.

\paragraph{Forward Preservation.}
As established in Equation~\eqref{eq:d2d-function-preservation}, the
replicated channels produce identical hidden activations at initialization,
and their new down-projection matrices sum to zero. Their combined
contribution therefore vanishes, while the original FFN parameters remain
unchanged. Consequently,
$\mathbf{F}_{\mathrm{Dense}}^{+}(\mathbf{x})
=\mathbf{F}_{\mathrm{Dense}}(\mathbf{x})$ for any input $\mathbf{x}$,
showing that Inherit4Rec-D2D achieves exact function preservation at
expansion in exact arithmetic, before further training.

\begin{figure}[!htbp]
\centering
\includegraphics[width=0.90\linewidth]{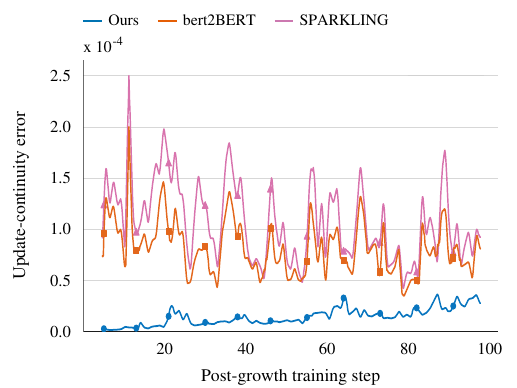}
\captionsetup{labelsep=colon,font=small,labelfont=bf,textfont=bf,skip=3pt}
\caption{D2D discrepancy $\mathcal E_t^{\mathrm u}$ between outputs
computed using the parameter blocks retained from the original FFN and
those of an unexpanded reference during post-growth training. Lower is better.}
\label{fig:d2d-update-continuity}
\Description{D2D update-continuity curves for Inherit4Rec, bert2BERT, and SPARKLING. Outputs computed using the parameter blocks retained from the original FFN are compared with an unexpanded reference as both models undergo further training. Inherit4Rec has lower output discrepancy.}
\end{figure}

\paragraph{Update Continuity.}
To assess the stability of the inherited channels' optimization
trajectories, we identify the gate, up, and down parameter blocks retained
from the original FFN, excluding all newly added blocks.
As these retained parameters are updated during post-growth training, we
compare the outputs computed using them with those of an unexpanded
reference trained further from the same checkpoint. We evaluate the
normalized RMS discrepancy $\mathcal{E}_t^{\mathrm{u}}$ on identical
inputs at matched training steps, using the source FFN's output RMS for
normalization. As shown in Figure~\ref{fig:d2d-update-continuity},
Inherit4Rec-D2D exhibits lower discrepancies and smaller fluctuations than
bert2BERT-FPI~\citep{chen2021bert2bert} and SPARKLING~\citep{yu2026sparkling}.
The smaller fluctuations provide indirect evidence of more stable update
trajectories for the inherited parameter blocks during continued training,
supporting the benefit of preserving their optimization history during
model growth.

\begin{figure}[!htbp]
\centering
\includegraphics[width=0.90\linewidth]{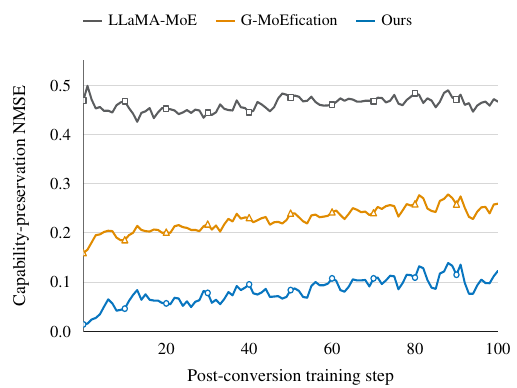}
\captionsetup{labelsep=colon,font=small,labelfont=bf,textfont=bf,skip=3pt}
\caption{Normalized MSE $\mathcal E_t^{\mathrm c}$ of D2S outputs against a Dense reference at matched activation budgets. Lower is better.}
\label{fig:d2s-capability-preservation}
\Description{D2S capability-preservation curves for Inherit4Rec, LLaMA-MoE, and G-MoEfication; Inherit4Rec has lower normalized MSE of FFN outputs.}
\end{figure}

\paragraph{Capability Preservation.}
Starting from the same Dense checkpoint, we further train each converted
model and an unconverted Dense reference on the same data. The converted
models use matched FFN activation budgets, and we evaluate the normalized
MSE $\mathcal{E}_t^{\mathrm{c}}$ of their outputs against the Dense reference
on identical inputs at matched training steps. As shown in
Figure~\ref{fig:d2s-capability-preservation}, Inherit4Rec-D2S maintains
lower output errors than LLaMA-MoE~\citep{zhu2024llama} and
G-MoEfication~\citep{lee2024breaking} throughout the evaluated training
period. This indicates that Inherit4Rec-D2S better preserves the Dense
FFN's outputs under sparse activation, with the advantage persisting as
training progresses. These results suggest that retaining high-contribution
channels in a shared expert and grouping co-active channels help preserve
the source model's learned capabilities under a limited activation budget.

\paragraph{Balanced Activation.}
By introducing a Switch-style load-balancing loss~\citep{fedus2021switch},
Inherit4Rec-D2S achieves balanced utilization of routed experts during
further training. Expert-load visualizations are provided in
Figure~\ref{fig:d2s-expert-load} in
Appendix~\ref{app:additional-evaluation-metrics}.

% Finish the diagnostic figures before balancing the remaining main-paper text.
\FloatBarrier
\balance
\endgroup

%% file: section/table_d2d.tex
% !TEX root = ../main.tex
\begin{table*}
\caption{Offline Dense-to-Dense GAUC results. Bold and underlined values
indicate the best and second-best results among transformation methods,
respectively.}
\label{tab:d2d-main}
\centering
\footnotesize
\setlength{\tabcolsep}{1.1pt}
\renewcommand{\arraystretch}{1.05}
\begin{threeparttable}
\begin{tabular*}{\linewidth}{
@{\extracolsep{\fill}}
P{0.20\linewidth}
*{8}{c}
@{}}
\toprule
& \multicolumn{4}{c}{KuaiRand-1K Dataset}
& \multicolumn{4}{c}{Industrial Dataset} \\
\cmidrule(lr){2-5}
\cmidrule(lr){6-9}

Model
& Effective-view & Long-view & Comment & Like
& Effective-view & Long-view & Follow & Like \\
\midrule

Seed (Base)
& 0.6631 & 0.6987 & 0.6773 & 0.6690
& 0.7545 & 0.7655 & 0.8259 & 0.8293 \\

Train-from-Scratch (Oracle)
& 0.6659 & 0.7013 & 0.6935 & 0.6707
& 0.7557 & 0.7667 & 0.8273 & 0.8307 \\

\midrule

bert2BERT-FPI
& 0.6638 & 0.6995 & 0.6857 & 0.6661
& \underline{0.7548} & \underline{0.7658}
& \underline{0.8266} & \underline{0.8298} \\

SPARKLING
& \underline{0.6653} & \underline{0.6998}
& \underline{0.6904} & \underline{0.6691}
& 0.7547 & 0.7657 & 0.8265 & 0.8297 \\

\midrule

Inherit4Rec-D2D
& \textbf{0.6656} & \textbf{0.7010}
& \textbf{0.6914} & \textbf{0.6706}
& \textbf{0.7551} & \textbf{0.7662}
& \textbf{0.8272} & \textbf{0.8302} \\

Impr.
& 0.05\% & 0.17\% & 0.14\% & 0.22\%
& 0.04\% & 0.05\% & 0.07\% & 0.05\% \\

\bottomrule
\end{tabular*}

\begin{tablenotes}[flushleft]
\footnotesize
\item Seed and Train-from-Scratch are reference models and are excluded
from ranking and baseline selection. Impr. is the relative GAUC gain of
Inherit4Rec-D2D over the strongest competing transformation method:
$100\times(\mathrm{GAUC}_{\mathrm{Ours}}
-\mathrm{GAUC}_{\mathrm{base}})
 /\mathrm{GAUC}_{\mathrm{base}}$.
\end{tablenotes}
\end{threeparttable}
\end{table*}

%% file: section/conclusion.tex
% !TEX root = ../main.tex
\section{Conclusion}
\label{sec:conclusion}

We present Inherit4Rec, a parameter-inheritance framework for evolving recommender models during streaming training. Its Dense-to-Dense method expands per-token FFNs while preserving the source function at conversion and transferring optimizer states; its Dense-to-Sparse method uses channel contribution and co-activation statistics to reorganize existing channels into shared and routed experts under a controlled active-width budget. On KuaiRand-1K and an industrial short-video recommendation stream, both methods outperform the evaluated checkpoint-based transformation baselines across prediction objectives. These results show that model capacity can be expanded or sparsified while reusing accumulated training state.

%% file: section/appendix.tex
% !TEX root = ../main.tex

\section{Additional Methodological Details}
\label{app:method-details}

This section supplements Section~\ref{sec:method} with the statistics,
partition solver, and state mappings needed to construct a transformed
checkpoint. The D2D initialization and D2S training objectives follow the
main text; the details below specify how channel allocation and optimizer
inheritance are carried out before continued training.

\subsection{Dense-to-Sparse Calibration and Partitioning}
\label{app:d2s-partition-algorithm}

\paragraph{Calibration statistics.}
We freeze the Dense checkpoint and collect $N$ inputs for each transformed
FFN from pre-conversion training examples. The $m$ channels with the largest
contribution energies in Equation~\eqref{eq:d2s-contribution-energy} form
the shared expert; $U$ contains the remaining $Rm$ channels.
The shared and routed channel sets are selected independently for each
FFN, so modules need not share a channel partition. Calibration uses
forward passes through the frozen source model rather than parameter
updates to that model.
For each input, let
$a_j(\mathbf x)=|h_j(\mathbf x)|\|W_{\mathrm{down},j:}\|_2$
be channel $j$'s output-contribution magnitude. Among channels in $U$, we
retain the top-$K_c$ contributions exceeding $\gamma$:
\begin{equation}
\begin{aligned}
b_j(\mathbf x_n)=\mathbb I\bigl[
&j\in\operatorname{Top}_{K_c}\{a_l(\mathbf x_n):l\in U\}\\
&{}\land a_j(\mathbf x_n)>\gamma\bigr].
\end{aligned}
\label{eq:d2s-contribution-indicator}
\end{equation}
Using output contributions accounts for both hidden activations and the
associated down-projection weights. The top-$K_c$ operation is applied
after removing the shared channels, so the co-activation statistics
describe the channels that still need to be grouped into routed experts.
We accumulate pairwise co-activation and marginal support counts,
\begin{equation}
\begin{aligned}
C_{ij}&=\sum_{n=1}^{N}b_i(\mathbf x_n)b_j(\mathbf x_n),\\
n_i&=\sum_{n=1}^{N}b_i(\mathbf x_n),
\end{aligned}
\label{eq:d2s-coactivation-count}
\end{equation}
independently for each FFN.

\paragraph{Distributed collection.}
Let $S_j^{(w)}$ be worker $w$'s sum of squared contributions for channel
$j$, and $N^{(w)}$ its input count. Worker-local statistics are combined as
\begin{equation}
\begin{aligned}
I_j&=\frac{\sum_w S_j^{(w)}}{\sum_w N^{(w)}},\\
C_{ij}&=\sum_w C_{ij}^{(w)},\qquad
n_i=\sum_w n_i^{(w)}.
\end{aligned}
\label{eq:d2s-distributed-statistics}
\end{equation}
Contribution energy is a mean over inputs, whereas co-activation and
support remain counts. Keeping the latter as counts preserves the
support scale used by the shrinkage term below.

\paragraph{Support-aware similarity.}
The support-aware similarity is
\begin{equation}
G_{ij}=\frac{C_{ij}}{\sqrt{n_i n_j}}
       \frac{C_{ij}}{C_{ij}+\tau_{\mathrm s}},
\label{eq:d2s-coactivation-similarity}
\end{equation}
with $G_{ij}=0$ when the normalization denominator is zero.
The shrinkage parameter $\tau_{\mathrm s}$ reduces the influence of
channel pairs supported by few observations.
The first factor normalizes co-activation by the channels' marginal
frequencies; the second discounts apparently strong associations based
on limited support. Thus, similarity reflects how consistently two
channels contribute together, rather than their co-occurrence count alone.

\paragraph{Partition solver.}
We approximate the equal-width partition objective in
Equation~\eqref{eq:d2s-allocation-objective} as follows.
\begin{enumerate}[leftmargin=*,itemsep=2pt]
\item \textbf{Select seeds.}
Compute weighted degrees $d_i=\sum_{j\in U}G_{ij}$ and use the
highest-degree channel as the first seed. For each subsequent seed,
retain the highest-degree $\min(|\mathcal C|,\max(256,64R))$ unused
candidates, where $\mathcal C$ is the unused candidate set. Select the
channel maximizing
\begin{equation}
q_i=\frac{d_i}{d_{\max}}
-\max_{s\in\mathcal Z}
\frac{G_{is}}{\sqrt{(d_i+\epsilon)(d_s+\epsilon)}},
\label{eq:d2s-seed-score}
\end{equation}
where $\mathcal Z$ contains the selected seeds, $d_{\max}$ is the largest
candidate degree, and $\epsilon>0$ stabilizes the similarity term.
The degree term favors channels with strong connections to the remaining
graph, while the similarity penalty discourages selecting redundant
seeds for different experts.
\item \textbf{Assign channels.}
Process the remaining channels in descending degree order and assign each
channel $i$ to the expert maximizing
\begin{equation}
A(i,r)=\frac{\sum_{j\in E_r}G_{ij}}{\sqrt{|E_r|}}
       +\beta(m-|E_r|),
\label{eq:d2s-assignment-score}
\end{equation}
restricted to experts with $|E_r|<m$.
The affinity term measures compatibility with the channels already in an
expert. The capacity bonus favors experts with more available slots,
while the hard capacity constraint ensures that every routed expert ends
with exactly $m$ channels.
\item \textbf{Refine and select.}
For each expert pair, apply the channel swap giving the largest positive
increase in the partition objective. Swaps preserve expert widths.
Run at most two refinement rounds, stopping early if a round yields no
positive-gain swap. Generate four candidates by randomly breaking ties
among equal-degree channels, then select the refined partition with the
largest objective value. Together with the shared expert, it covers all
source FFN channels without overlap.
\end{enumerate}
The search is a heuristic for the objective in the main text; it does
not assume a globally optimal partition. Equal channel counts determine
expert capacity at construction, while the separate balancing loss
controls how often experts are selected during continued training.

\subsection{State Migration and Continued Training}
\label{app:state-migration}

\paragraph{Dense-to-Dense growth.}
Inherited parameters retain their original indices, optimizer moments,
and learning-rate schedule. New parameters start with zero moments and
follow the separate warmup-and-decay schedules in
Table~\ref{tab:method-hyperparameters}. This includes the new gate and up
projections, whose parameter values are copied from the source FFN but
whose optimizer moments are initialized independently.
In particular, the first- and second-order moments of an inherited block
remain associated with that same block after expansion. A copied input
projection is a new parameter block even when its initial value equals
the source block. This distinction preserves the optimization history of
the existing channels while allowing the added channels to follow their
own learning-rate schedule.

The dataset-specific growth factors instantiate the initialization in
Equation~\eqref{eq:d2d-group-centering}. The industrial setting adds three
replica blocks ($c=3$); KuaiRand-1K adds two ($c=2$). Each candidate
down-projection block uses the global entrywise mean and variance of the
source down projection. Centering across the two new blocks in the latter
case gives a pair $A$ and $-A$. In both settings, zero-sum centering is
applied only at initialization, and the new blocks subsequently train
without this constraint.

\paragraph{Dense-to-Sparse conversion.}
For an expert with channel set $B$, parameters and their corresponding
optimizer moments use the same channel-aligned slices:
\begin{equation}
\begin{aligned}
W_{\mathrm{gate}}^{(B)}&=W_{\mathrm{gate},:,B},\\
W_{\mathrm{up}}^{(B)}&=W_{\mathrm{up},:,B},\\
W_{\mathrm{down}}^{(B)}&=W_{\mathrm{down},B,:}.
\end{aligned}
\label{eq:d2s-parameter-slicing}
\end{equation}
Parameters and optimizer states outside the converted FFNs remain
unchanged at conversion. The bias-free router uses a zero-mean truncated
normal initialization with standard-deviation parameter $0.01$ and fresh
optimizer states. All shared and routed experts remain trainable; routing
and the training loss follow Equations~\eqref{eq:d2s-sparse-output}
and~\eqref{eq:d2s-training-objective}.
The same channel indices must be used for all three projections and their
moment tensors: selecting columns of the gate and up projections requires
selecting the corresponding rows of the down projection. This mapping
retains the source channels' parameters and accumulated moments while
changing which channels are executed together. It does not make sparse
execution functionally identical to the Dense FFN, because each input
uses only the shared expert and one routed expert.

The shared-plus-top-1 rule applies during both continued training and
inference. The auxiliary balancing losses are averaged over transformed
FFNs before being combined with the task loss. Experts are therefore
initialized by inheritance but continue adapting along with the new
router on subsequent training data.

\section{Design Principles and Diagnostic Metrics}
\label{app:additional-evaluation-metrics}

The following definitions formalize the four design principles evaluated
in Section~\ref{sec:exp-additional-analyses}.
Forward preservation concerns the instant of D2D growth;
update continuity and capability preservation evaluate behavior during
continuation. Balanced activation concerns the aggregate routing load.

\begin{definition}[Forward Preservation]
An expanded FFN preserves the source function if
$F_{\boldsymbol{\theta}_t^{\mathrm d}(0)}(\mathbf x)
=F_{\boldsymbol{\theta}_t}(\mathbf x)$ for every input immediately after
growth, in exact arithmetic. Equation~\eqref{eq:d2d-function-preservation}
establishes this property for Inherit4Rec-D2D.
\end{definition}

\begin{definition}[Update Continuity]
Update continuity requires smooth and stable optimization trajectories
for the inherited channels during subsequent training.
Let $\boldsymbol{\theta}_{t,\mathrm{old}}^{\mathrm d}(\tau)$ denote the
inherited parameter blocks at continuation step $\tau$, and
$\bar{\boldsymbol{\theta}}_t(\tau)$ an unexpanded reference continued
from the same checkpoint. To examine update continuity empirically, we
use their normalized RMS output discrepancy:
\begin{equation}
\mathcal E_t^{\mathrm u}(\tau,\mathbf x)=
\frac{
\operatorname{RMS}\!\left(
F_{\boldsymbol{\theta}_{t,\mathrm{old}}^{\mathrm d}(\tau)}(\mathbf x)
-F_{\bar{\boldsymbol{\theta}}_t(\tau)}(\mathbf x)
\right)
}{
\operatorname{RMS}\!\left(F_{\boldsymbol{\theta}_t}(\mathbf x)\right)
+\epsilon
}.
\end{equation}
The denominator uses the source FFN before growth on the same input.
Only the retained gate, up, and down blocks contribute to the expanded
model's output in this diagnostic; newly added blocks are excluded.
The temporal fluctuations of this discrepancy provide indirect evidence
about the stability of the inherited channels' optimization trajectories.
Smaller fluctuations are consistent with more stable updates during
continued training. This output-based diagnostic does not directly
measure parameter-update stability.
\end{definition}

\begin{definition}[Capability Preservation]
Let $\boldsymbol{\theta}_t^{\mathrm m}(\tau)$ denote a converted MoE and
$\bar{\boldsymbol{\theta}}_t(\tau)$ an unconverted Dense reference
continued from the same checkpoint. At matched continuation steps, their
output NMSE is
\begin{equation}
\mathcal E_t^{\mathrm c}(\tau)=
\frac{
\mathbb E_{\mathbf x\sim P_t(\mathbf X)}
\left[\left\|
F_{\boldsymbol{\theta}_t^{\mathrm m}(\tau)}(\mathbf x)
-F_{\bar{\boldsymbol{\theta}}_t(\tau)}(\mathbf x)
\right\|_2^2\right]
}{
\mathbb E_{\mathbf x\sim P_t(\mathbf X)}
\left[\left\|F_{\bar{\boldsymbol{\theta}}_t(\tau)}(\mathbf x)
\right\|_2^2\right]+\epsilon
}.
\end{equation}
The MoE activates the shared expert and one top-1 routed expert. At
$\tau=0$, the reference is the pre-conversion FFN. Lower NMSE indicates
closer preservation of the reference output behavior.
For $\tau>0$, the reference is also trained further. The comparison
therefore measures sparse-model fidelity to a Dense model exposed to the
same continuation data, rather than fidelity to a frozen checkpoint.
The normalized output error supplements the downstream GAUC evaluation;
it is not itself a prediction-quality metric.
\end{definition}

\begin{definition}[Balanced Activation]
Let $f_r\geq0$ be routed expert $r$'s hard-assignment fraction, with
$\sum_{r=1}^{R}f_r=1$. Balanced activation corresponds to fractions near
$1/R$. The always-on shared expert is excluded from this calculation.
\end{definition}
Equal expert widths and balanced activation are distinct properties.
The partition solver fixes the number of channels in each expert, but
the router determines the sample fractions $f_r$. The auxiliary loss in
Equation~\eqref{eq:d2s-balance-loss} encourages balanced use of these
equal-capacity experts during training.

\begin{figure}[t]
\centering
\includegraphics[width=\columnwidth,trim=0 12bp 0 0,clip]{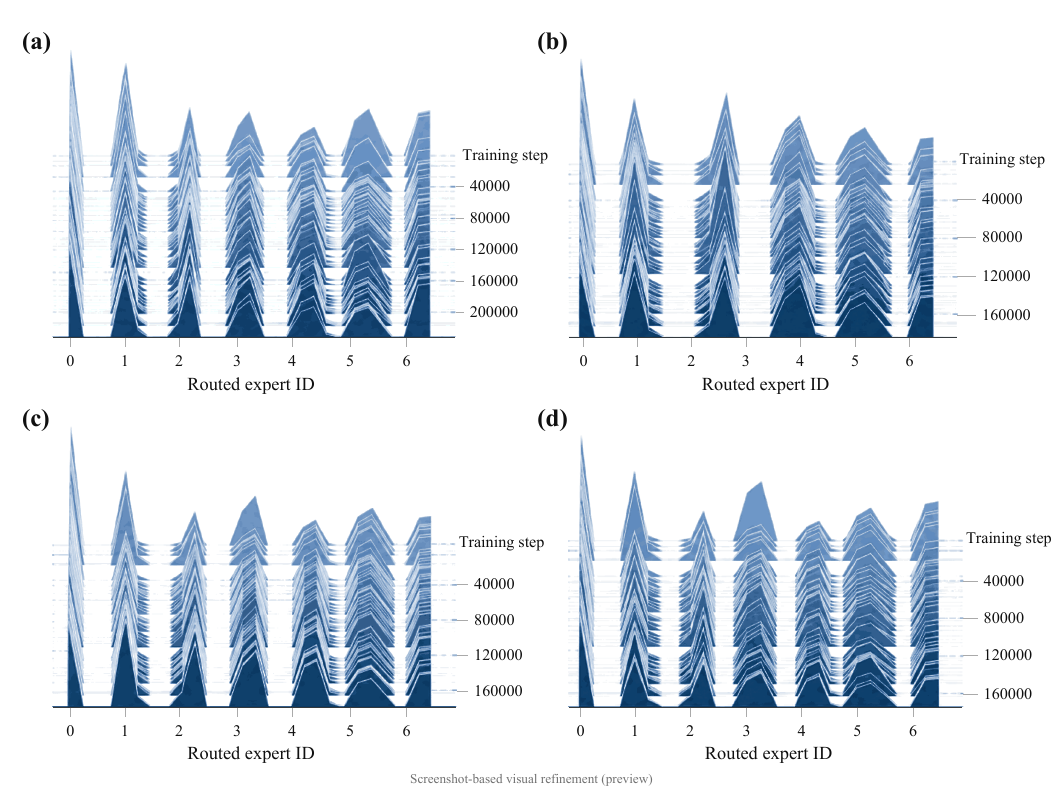}
\caption{Examples of routed-expert load distributions during continued
training. Each panel shows seven routed experts, with stacked profiles
corresponding to successive training steps. The shared expert is excluded.}
\label{fig:d2s-expert-load}
\Description{Four panels show sampled load profiles for routed experts
zero through six at successive training steps. All routed experts remain
active, with load variations over time.}
\end{figure}

\paragraph{Diagnostic evaluation.}
Transformed models and references use the same continuation data; FFN
outputs are evaluated on identical inputs at matched training steps.
D2S activation budgets are matched across converted models.
Figure~\ref{fig:d2s-expert-load} illustrates routed-expert utilization
during training with the Switch-style balancing loss~\citep{fedus2021switch}.
Figures~\ref{fig:d2d-update-continuity}
and~\ref{fig:d2s-capability-preservation} in the main text use the RMS
discrepancy and NMSE defined above, respectively. The D2D denominator is
the pre-growth source output, whereas the D2S denominator is the Dense
reference output at the matched continuation step. The two diagnostics
therefore use different normalizers and are not on a directly comparable
numerical scale.

\paragraph{Expert-load profiles.}
The four panels in Figure~\ref{fig:d2s-expert-load} show sampled profiles
for seven routed experts. Loads vary
across training steps, but the displayed profiles show activity across
the expert set rather than sustained concentration on one expert.

\section{Additional Experimental Details and Results}
\label{app:experimental-details}
\label{app:additional-analyses}

\subsection{Datasets and Training Protocol}
\label{app:datasets-and-evaluation}

\paragraph{Datasets and prediction tasks.}
Each example is a user--video interaction recording the response to an
exposed video. KuaiRand-1K includes Effective-view, Long-view, Comment,
and Like; the industrial dataset includes Effective-view, Long-view,
Follow, and Like. The industrial data come from a single page of a
short-video recommendation platform. Tables~\ref{tab:d2d-main}
and~\ref{tab:d2s-main} report GAUC separately for each task.

\paragraph{Industrial stream.}
For D2D, source models train on the first 18 daily traffic windows,
followed by seven continuation windows after growth; the
Train-from-Scratch reference uses the complete 25-window stream.
For D2S, source models train on the first 20 daily traffic windows,
followed by eight continuation windows after conversion; the
Train-from-Scratch references use the complete 28-window stream.
Each daily window contains approximately 50 billion examples.
In D2S, the small Dense reference
Seed (Base) uses the same eight continuation windows as the converted
models and is evaluated after this continuation. D2S calibration examples
are sampled from the pre-conversion stream.
This chronological protocol separates data used to obtain the source
checkpoint from the subsequent stream used to evaluate adaptation after
transformation. The source checkpoint and continuation windows are shared
among inheritance methods within each comparison, so their differences are
evaluated from the same starting model and subsequent data.

\paragraph{KuaiRand-1K}
D2D uses five source-model epochs and five post-growth epochs; D2S uses
three complete epochs after conversion. The D2S Seed (Base) also completes
three epochs of continued training on the same data before evaluation.
Calibration uses pre-conversion
training examples. Sparse embedding tables and their SparseAdam states
are reinitialized between non-final epochs, while dense parameters and
AdamW states are retained. This epoch-level reset is distinct from state
migration at model transformation.
The calibration set contains 512 examples in eight batches of 64. This
is the number of calibration examples, while $N$ in the methodological
definitions denotes inputs to an individual FFN. Epoch-level embedding
resets should also be distinguished from the optimizer-transfer ablation,
which concerns the training state retained at the transformation boundary.

\subsection{Comparison Protocol and Configurations}
\label{app:method-hyperparameters}

\paragraph{Comparison controls and references.}
The D2D inheritance baselines are bert2BERT-FPI and SPARKLING; the D2S
baselines are LLaMA-MoE and G-MoEfication, as in Section~\ref{sec:exp-setup}.
Competing inheritance methods within each comparison share the source
checkpoint and continuation data. D2S transformation methods also share
calibration examples and FFN activation budgets.

For D2D, Seed (Base) continues training the source model without expansion,
using the same continuation data and horizon as the expanded models.
Train-from-Scratch (Oracle) provides the target dense reference.

For D2S, the source checkpoint is a pretrained large Dense model.
Seed (Base) is a separate small Dense reference whose active parameter
count is comparable to that of the converted sparse models. On both
datasets, it undergoes continued training on the same data and for the
same duration as the converted models. Dense Train-from-Scratch provides
a large dense reference, while SMoE Train-from-Scratch (Oracle) matches
the target sparse architecture. These reference models are excluded when selecting the
strongest inheritance baseline for the relative improvements in the
main result tables.

\paragraph{Configuration and parameter accounting.}
Table~\ref{tab:method-hyperparameters} summarizes dataset-specific settings.
Both datasets use $K_c=32$ and $\lambda_{\mathrm{bal}}=0.01$. On the
industrial dataset, $\gamma=0$, $\tau_{\mathrm s}=10$, and $\beta=10^{-7}$.

Parameter counts exclude sparse embedding tables; D2S counts include
routers. Active FFN width is $2m$ (one shared and one routed expert).
For D2S, the converted model's active parameter count is comparable to
that of Seed (Base), while its expert parameters are inherited from
the larger Dense source model.
The active-width ratios in the table describe FFN computation budgets,
not whole-model parameter or FLOPs reductions.
For the industrial setting, a width-1,024 FFN becomes one shared and
seven routed experts of width 128, with 256 channels active per input.
For KuaiRand-1K, a width-1,800 FFN becomes one shared and five routed
experts of width 300, with 600 channels active. The total channel count
is preserved in both cases. Router parameters explain why total D2S
parameter counts need not be identical to those of the source Dense
model even though its FFN channels are partitioned without duplication.

\begin{table*}[t]
\caption{Model and continued-training configurations. Parameter counts exclude sparse embedding tables; D2S counts include routers.}
\label{tab:method-hyperparameters}
\label{tab:kuairand-method-hyperparameters}
\centering
\footnotesize
\renewcommand{\arraystretch}{1.08}
\begin{tabularx}{\linewidth}{@{}P{0.26\linewidth}YY@{}}
\toprule
Setting & Industrial dataset & KuaiRand-1K \\
\midrule
Backbone & Three Transformer layers; hidden size 256; four heads; embedding size 64 & Two FIM layers and one TIM layer; hidden size 400; five heads; embedding size 128; 66 FIM tokens \\
Dense optimizer & AdamW; learning rate $2\times10^{-3}$; weight decay $10^{-3}$ & AdamW; learning rate $10^{-4}$; weight decay $10^{-3}$ \\
Sparse-embedding optimizer & Adam; learning rate $10^{-3}$ & SparseAdam; learning rate $10^{-3}$ \\
\midrule
D2D parameters & $28\,\mathrm M\rightarrow74\,\mathrm M$ & $112\,\mathrm M\rightarrow302\,\mathrm M$ \\
D2D FFN width & $256\rightarrow1{,}024$; $c=3$ & $600\rightarrow1{,}800$; $c=2$ \\
New-parameter warmup & 500 steps to $3\times10^{-3}$ & 200 steps from zero to $1.5\times10^{-4}$ \\
Subsequent cosine annealing & 1,000 steps to $2\times10^{-3}$ & 1,000 steps to $10^{-4}$ \\
\midrule
D2S total FFN width & $1{,}024$ & $1{,}800$ \\
Routed experts and expert width & $R=7$; $m=128$ & $R=5$; $m=300$ \\
Active FFN width (ratio) & $256$ ($1/4$) & $600$ ($1/3$) \\
D2S total/active parameters & $74\,\mathrm M/28\,\mathrm M$ & $303\,\mathrm M/113\,\mathrm M$ \\
Calibration examples & Approximately $7.68\times10^8$ & 512 (eight batches of 64) \\
\bottomrule
\end{tabularx}
\end{table*}

% Queue both configuration and compatibility tables before the final analysis.
\begin{table*}[t]
\caption{Backbone compatibility on RankMixer: offline GAUC of Seed (Base), Train-from-Scratch (Oracle), and Inherit4Rec-D2D and Inherit4Rec-D2S models on both datasets.}
\label{tab:rankmixer-compatibility}
\centering
\footnotesize
\setlength{\tabcolsep}{1.1pt}
\renewcommand{\arraystretch}{1.05}
\begin{tabular*}{\linewidth}{
@{\extracolsep{\fill}}
P{0.20\linewidth}
*{8}{c}
@{}}
\toprule
& \multicolumn{4}{c}{KuaiRand-1K Dataset}
& \multicolumn{4}{c}{Industrial Dataset} \\
\cmidrule(lr){2-5}
\cmidrule(lr){6-9}
Model
& Effective-view & Long-view & Comment & Like
& Effective-view & Long-view & Follow & Like \\
\midrule
Seed (Base)
& 0.6475 & 0.6827 & 0.6434 & 0.6642 & 0.7521 & 0.7337 & 0.8100 & 0.8216 \\
Train-from-Scratch (Oracle)
& 0.6496 & 0.6872 & 0.6601 & 0.6657 & 0.7531 & 0.7347 & 0.8127 & 0.8230 \\
\midrule
Inherit4Rec-D2D
& 0.6491 & 0.6866 & 0.6581 & 0.6650 & 0.7528 & 0.7343 & 0.8116 & 0.8225 \\
Inherit4Rec-D2S
& 0.6490 & 0.6861 & 0.6574 & 0.6649 & 0.7525 & 0.7341 & 0.8108 & 0.8220 \\
\bottomrule
\end{tabular*}
\end{table*}

\subsection{Computational Cost Accounting}
\label{app:cost-accounting}

For a cost measure $C$, savings relative to the target Train-from-Scratch
reference are $100(1-C_{\mathrm{inherit}}/C_{\mathrm{reference}})\%$.
The reported 72\% saving measures the industrial D2D incremental training
budget in daily windows. Starting from an existing Dense checkpoint
trained for 18 windows, the target model continues training for seven
windows, compared with 25 windows from scratch:
\begin{equation}
\mathrm{Savings}_{\mathrm{D2D}}
=100\left(1-\frac{7}{25}\right)\%=72\%.
\label{eq:d2d-incremental-training-savings}
\end{equation}
This training-window budget excludes the sunk cost of source pretraining
and the one-time growth overhead.

D2S incremental cost covers conversion and continuation from an available
Dense checkpoint, excluding source pretraining. Its industrial training
horizons are eight windows for continuation and 28 from scratch.
Active-width ratios describe FFN budgets; GPU-time savings also depend on
execution overhead and the remaining model components.

% Balance the final appendix page beneath the two wide tables.
\balance
\subsection{Backbone Compatibility}
\label{app:backbone-compatibility}

Table~\ref{tab:rankmixer-compatibility} reports GAUC with
RankMixer~\citep{rankmixer}. D2D grows from the small Dense checkpoint,
while Seed (Base) continues training that checkpoint without expansion.
D2S converts a separate pretrained large Dense checkpoint to an MoE
with approximately the same total parameter count and an active count
comparable to that of Seed (Base). Within each comparison, Seed (Base)
and the transformed models use the same continuation data and duration.
This experiment
assesses backbone compatibility without competing transformation methods.
Both transformations improve on Seed (Base) across all four tasks on
both datasets, while remaining below Train-from-Scratch (Oracle).
D2D evaluates whether growth can reuse a smaller checkpoint
with a different backbone; D2S evaluates whether the large backbone's
parameters can support sparse execution at the smaller active budget.

\subsection{Extended and Large-Scale Growth}
\label{app:extended-growth}

\paragraph{Extended post-growth training.}
Figure~\ref{fig:d2d-extended-training}(a) tracks the Effective-view GAUC
difference between the expanded model and Train-from-Scratch (Oracle)
as additional industrial streaming data are used after growth. This
experiment examines whether the gap to the larger from-scratch model
narrows beyond the main continuation comparison. Its reference is the
large target model, and the zero line denotes equal GAUC with that
reference.

\paragraph{Growth to one billion parameters.}
For Figure~\ref{fig:d2d-extended-training}(b), a 0.31B-parameter model
pretrained for four months is expanded to 1B parameters. Both the expanded
model and the original 0.31B model then continue training for 45 days.
At day 45, the expanded model achieves an absolute Effective-view GAUC
gain of 0.14\% over continued training of the original model.
Here the reference is the unexpanded 0.31B model, so the measured gain
assesses growth relative to continued use of the existing architecture.
The two panels use different references and therefore address distinct
questions: approaching a larger from-scratch target and improving on a
mature checkpoint through expansion.